\documentstyle[preprint,eqsecnum,aps,psfig]{revtex}
\begin{document}
\draft
\title{Quantum Chaos in Compact Lattice QED}
\author{B.A.~Berg}
\address{Department of Physics, The Florida State University,
Tallahassee, FL 32306, USA\\
Supercomputer Computations Research Institute, The Florida State University,\\
Tallahassee, FL 32306, USA}
\author{H.~Markum}
\address{Institut f\"{u}r Kernphysik, Technische Universit\"{a}t Wien,
 A-1040 Vienna, Austria}
\author{R.~Pullirsch}
\address{Department of Physics, The Florida State University, Tallahassee, 
FL 32306, USA\\
Institut f\"{u}r Kernphysik, Technische Universit\"{a}t Wien, A-1040 Vienna,
 Austria}
\date{\today}
\maketitle
\begin{abstract}
Complete eigenvalue spectra of the staggered Dirac operator in
quenched $4d$ compact QED are studied on $8^3 \times 4$ and $8^3 \times 6$
lattices. We investigate the behavior of the nearest-neighbor
spacing distribution $P(s)$ as a measure of the fluctuation
properties of the eigenvalues in the strong coupling and the
Coulomb phase. In both phases we find agreement with the Wigner
surmise of the unitary ensemble of random-matrix theory
indicating quantum chaos. Combining this with previous 
results on QCD, we conjecture that quite
generally the non-linear couplings of quantum field theories 
lead to a chaotic behavior of the eigenvalues of the Dirac
operator.
\end{abstract}

% body of paper here
\narrowtext

%%%%%%%%%%%%%%%%%%%%%%%%%%%%%%%%%%%%%%%%%%%%%%%%%%%%%%%%%%%%%%%%%%%%%%%%%%%%
\section{Motivation} \label{intro}

The fluctuation properties of the eigenvalues of the Euclidean lattice QCD Dirac
operator have attracted much attention in the past few years. In 
Ref.~\cite{Verb95} it was first shown for SU(2) lattice gauge theory 
that certain features of the spectrum of the Dirac operator are described
by random-matrix theory (RMT). In particular the
so-called nearest-neighbor spacing distribution $P(s)$, i.e. the distribution
of the spacings $s$ of adjacent eigenvalues on the ``unfolded'' scale (see
below), agrees with the Wigner surmise of RMT. According to the 
Bohigas-Giannoni-Schmit conjecture \cite{Bohi84b}, quantum systems whose
classical counterparts are chaotic have a $P(s)$ given by RMT whereas systems
whose classical counterparts are integrable obey a Poisson distribution
$P(s)=e^{-s}$. Therefore, the specific form of $P(s)$ is often taken as a
criterion for ``quantum chaos''. However, there is no accepted proof of the
Bohigas-Giannoni-Schmit conjecture yet. The field of quantum chaos is
still developing and there are many open conceptual problems \cite{riew}.
Applying this conjecture it was recently demonstrated that QCD is chaotic, both
in the confinement and the quark gluon plasma phase \cite{Pull98}.
 
A number of interesting results have been established for
chaotic dynamics in classical gauge theories. Lattice gauge
theories are chaotic as classical Hamiltonian dynamical systems 
\cite{Biro94}. Furthermore, it was found that the leading Lyapunov exponent
of SU(2) Yang-Mills field configurations indicates that configurations
corresponding to
the deconfinement phase are chaotic although they are less chaotic
than in the strong coupling phase at finite temperature \cite{Biro98}.
The scaling of the maximal Lyapunov exponent in the classical 
continuum limit was studied in Ref.~\cite{Gatto98}: It was
suggested that Abelian gauge theories behave regularly in the continuum limit
whereas non-Abelian gauge theories are chaotic in the continuum,
although the exact scaling relation is still an open problem.
Chaos to order transitions were observed in a spatially
homogeneous SU(2) Yang-Mills-Higgs system and in a spatially homogeneous
SU(2) Yang-Mills Chern-Simons Higgs system \cite{Sala97,Mukk97}.
In Ref.~\cite{Sala97} a chaos to order transition was also seen
on the quantum level, i.e. a smooth transition from a Wigner
to a Poisson distribution was found. 
A transition in $P(s)$ from Wigner to Poisson behavior was further
observed at the metal-insulator transition of the Anderson model
\cite{Alts88}.
Recently, the suppression of the characteristic manifestations
of dynamical chaos by quantum fluctuations was analyzed in 
the context of spatially homogeneous scalar electrodynamics \cite{Mati97}
and for a $0+1$-dimensional space-time $N$-component $\phi^4$ theory
in the presence of an external field \cite{Case98}.
These chaos to order transitions were seen in spatially
homogeneous models and not for the full classical field theory.
The relationship to properties of the quantum field theory is an
interesting issue.

Here we focus on the Dirac operator for quenched $4d$ compact QED
to search for the possible existence of a transition from chaotic to
regular behavior in Abelian lattice gauge
theories. In particular, we are interested in the 
nearest-neighbor spacing distribution of the eigenvalues of the Dirac 
operator across the phase transition from the strong coupling to the 
Coulomb phase. In the strong coupling region Abelian as well as non-Abelian 
lattice gauge theories are in a confined phase \cite{Wils74}. 
For compact QED this means that for couplings 
$\beta < \beta_c \approx 1.01$ the electron is confined.
However, when crossing the phase transition the conventional Coulomb phase is
observed. There are some interesting properties of the two phases
which can be studied in lattice QED. In the confinement phase the
photons form massive bound states similar as the gluons 
bind to glue-balls in lattice QCD.
When crossing the phase transition a massless photon is found \cite{Berg84}
whereas in lattice QCD the gluon is a massive particle in the deconfinement
region. U(1) lattice gauge theory contains Dirac magnetic monopoles in 
addition to photons \cite{Poly75} and it was demonstrated via numerical
simulations that the vacuum in the confined phase
is populated by monopole currents which become rare in the Coulomb phase
\cite{DeGr80}. It is an interesting question if the difference between
the Coulomb phase in QED and the quark-gluon plasma phase in QCD has an
influence on the level repulsion of the corresponding Dirac spectra. 

%%%%%%%%%%%%%%%%%%%%%%%%%%%%%%%%%%%%%%%%%%%%%%%%%%%%%%%%%%%%%%%%%%%%%%%%%%%%%
\section{Analysis} \label{sect2}

We generated gauge field configurations using the standard Wilson plaquette
action for U(1) gauge theory,
\begin{equation}
S_G(U_l)=\beta\sum\limits_{P}(1-\cos \Theta_P) \; ,
\end{equation}
where $U_l\equiv U_{x\mu}=\exp(i\Theta_{x\mu})$, with $\Theta_{x\mu}\in
[-\pi,\pi)$, are the field variables defined on the links $l\equiv(x,\mu)$.
The plaquette angles are $\Theta_P = \Theta_{x,\mu}+\Theta_{x+\hat{\mu},\nu}
-\Theta_{x+\hat{\nu},\mu}-\Theta_{x,\nu}$. We simulated $8^3 \times 4$
and $8^3 \times 6$ lattices at various values of the inverse gauge
coupling $\beta=1/e^2$
both in the strong coupling and the Coulomb phase. Typically we discarded
the first 10000 sweeps for reaching equilibrium and produced 20 
independent configurations separated by 1000 sweeps for 
each $\beta$. Because of the spectral ergodicity property of RMT one can
replace ensemble averages by spectral averages \cite{Guhr98} if one is
only interested in the bulk properties. Thus a few independent
configurations are sufficient to compute $P(s)$.

On the lattice the Dirac operator $/\!\!\!\!D=/\!\!\!\partial+ie\: /\!\!\!\!A$ 
for staggered fermions 
\begin{equation}
M_{x,x'}=\frac{1}{2}\sum\limits_{\mu=1}^{4}\eta_{x\mu} \left( 
\delta_{x+\hat{\mu},x'} U_{x,\mu} - \delta_{x-\hat{\mu},x'} 
U_{x,\mu}^{\dagger} \right) 
\end{equation}
is anti-Hermitian so that all 
eigenvalues are imaginary. For convenience we denote them by 
$i\lambda_n$ and refer to the $\lambda_n$ as the eigenvalues in the 
following. Because of $\{/\!\!\!\!D,\gamma_5\}=0$ the $\lambda_n$ 
occur in pairs of opposite sign.  All spectra were checked against 
the analytical sum rules
\begin{equation}
\sum_{n} \lambda_n = 0 \qquad {\rm and} \qquad
\sum_{\lambda_n>0} \lambda_n^2 = V \:,
\end{equation}
where $V$ is the lattice volume \cite{Tilo98}.
We further checked our spectra by calculating the chiral condensate
\begin{equation}
\langle \bar{\chi}\chi \rangle = V^{-1} \langle \sum_n \left(i\lambda_n
+m\right)^{-1} \rangle
\end{equation}
for $m=0.04$ and found agreement with results in the literature \cite{Hofe94}.

\begin{figure}[p]
  \centerline{\hbox{
  \psfig{figure=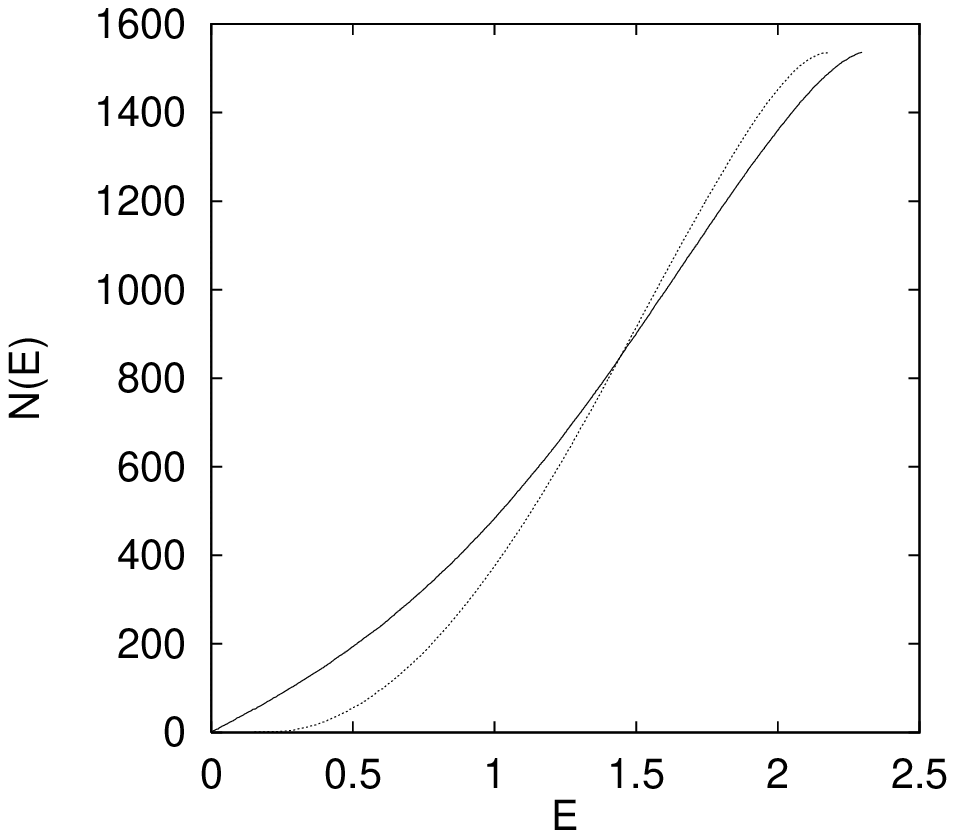,width=10cm}
  }}
  \caption{\label{fig1} Staircase function $N(E)$ representing the number
   of positive eigenvalues $\leq E$ for a typical configuration of compact
   U(1) theory on an
   $8^3 \times 6$ lattice in the strong coupling phase $\beta = 0.90$
   (solid line) and in the Coulomb phase $\beta = 1.10$ (dotted line).}
\end{figure}

To construct the nearest-neighbor spacing distribution $P(s)$ from 
the eigenvalues, one has to ``unfold'' the spectra. This procedure
is a local rescaling of the energy scale so that the mean level spacing
$\bar{s}$ is equal to unity on the unfolded scale \cite{Bohi84a}:
One first defines
the staircase function $N(E)$ to be the number of eigenvalues with
$\lambda\le E$. This staircase function is decomposed into an average
part and a fluctuating part, $N(E)=N_{\rm av}(E)+N_{\rm fl}(E)$. The
smooth average part is extracted by fitting $N(E)$ to a smooth curve,
e.g. to a low-order Chebyshev polynomial. One then defines the unfolded energies
to be $x_n=N_{\rm av}(E_n)$. As a consequence the sequence $\{x_n\}$ has
mean level spacing equal to unity. Ensemble and spectral averages are
only meaningful after unfolding. Figure~\ref{fig1} shows a typical
staircase function for $\beta = 0.90$ (strong coupling phase) and 
$\beta = 1.10$ (Coulomb phase) on an $8^3 \times 6$ lattice.
It exhibits a decrease of small eigenvalues due to the restoration
of chiral symmetry across the transition.

%%%%%%%%%%%%%%%%%%%%%%%%%%%%%%%%%%%%%%%%%%%%%%%%%%%%%%%%%%%%%%%%%%%%%%%%%%%%%
\section{Results and Discussion}

In RMT one has to distinguish between different universality classes
which are determined by the symmetries of the system. So far the 
classification for the QED Dirac operator has not been done. Our
calculations show that in the case of the staggered $4d$ compact QED Dirac matrix
the appropriate ensemble is the unitary ensemble. Although from a 
mathematical point of view this is the simplest one, the RMT result
for the nearest-neighbor spacing distribution is still rather complicated.
It can be expressed in terms of so-called prolate spheroidal functions,
see Ref.~\cite{Meht91} where $P(s)$ has also been tabulated. A good
approximation to $P(s)$ is provided by the Wigner surmise for the 
unitary ensemble
\begin{equation}
  \label{ue}
  P(s)=\frac{32}{\pi^2}\,s^2\,e^{-\frac{4}{\pi}s^2}\:.
\end{equation}

\begin{figure}[p]
  \centerline{\hbox{
  \psfig{figure=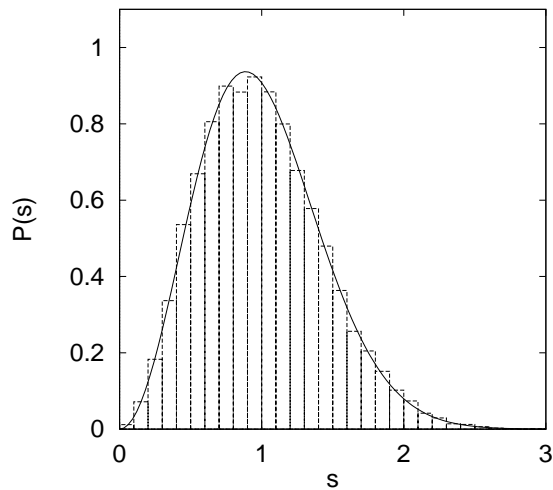,width=10cm}
  }}
  \caption{\label{fig2} Nearest-neighbor spacing distribution $P(s)$ 
       of the Dirac operator for compact U(1) theory in the strong
       coupling phase for $\beta=0.90$.
       The histogram represents the lattice data on an $8^3 \times 6$
       lattice averaged over 20 independent configurations. The full
       curve is the Wigner distribution of
       Eq.~(\protect\ref{ue}) for the unitary ensemble of RMT.}
\end{figure}

\begin{figure}[p]
  \centerline{\hbox{
  \psfig{figure=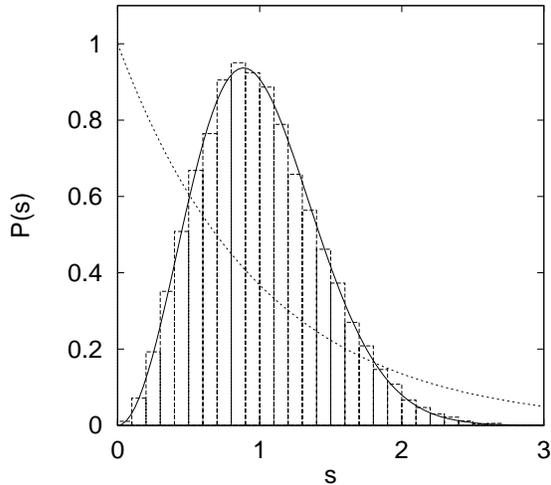,width=10cm}
  }}
  \caption{\label{fig3} Nearest-neighbor spacing distribution $P(s)$
       of the Dirac operator for compact U(1) theory in the Coulomb
       phase for $\beta=1.10$. The 
       histogram represents the lattice data on an $8^3 \times 6$
       lattice averaged over 20 independent configurations.
       The full curve is the Wigner distribution of
       Eq.~(\protect\ref{ue}) for the unitary ensemble of RMT.
       For comparison the Poisson distribution
       $P(s) = e^{-s}$ is also indicated by the dashed line.}
\end{figure}

We have simulated $8^3 \times 4$ lattices at $\beta=0,\:0.90,\:0.95,\:1.00,\:
1.05,\:1.10,\:1.50$ and $8^3 \times 6$ lattices at $\beta=0.90,\:1.10,\:
1.50$. All results are similar to those selected for the plots.
Figure~\ref{fig2} shows the 
nearest-neighbor spacing distribution $P(s)$ for $\beta=0.90$ in the
confined phase averaged over 20 independent configurations on the
$8^3 \times 6$ lattice compared with the Wigner surmise for the unitary
ensemble of RMT of Eq.~(\ref{ue}). Good agreement is found. According
to the Bohigas-Giannoni-Schmit conjecture this means the system can be
regarded as chaotic in the strong coupling region.
Figure~\ref{fig3} shows the nearest-neighbor spacing distribution 
$P(s)$ for $\beta=1.10$ in the Coulomb phase again averaged over 
20 independent configurations and compared
with the Wigner surmise (\ref{ue}).
The agreement of the lattice data with the RMT predictions is
interpreted as a signal that quantum chaos survives the phase transition.
We find no deviation up to the maximum coupling considered, $\beta = 1.50$. 

\begin{figure}[p]
  \centerline{\hbox{
  \psfig{figure=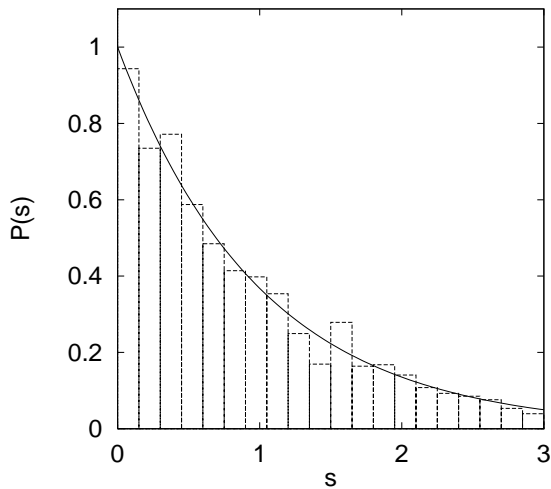,width=10cm}
  }}
  \caption{\label{fig4} Nearest-neighbor spacing distribution $P(s)$ 
       of the analytically calculated eigenvalues of
       Eq.~(\protect\ref{free}) for a free Dirac operator on a 
       $53\times 47 \times 43 \times 41$ lattice (histogram)
       compared with the Poisson distribution $P(s) = e^{-s}$
       (solid line).}
\end{figure}

In the strong coupling phase the result holds down to $\beta=0$.
Therefore, we tend to interpret our, as well as previous 
\cite{Pull98,Verb95}, results in the sense that the disorder of 
the gauge field configurations \cite{Biro94,Biro98} is responsible
for the chaotic characteristics of the spectrum of the Dirac operator.
In contrast 
to that: The free fermion theory is non-chaotic and the
corresponding
nearest-neighbor spacing distribution obeys a Poisson distribution.
This is illustrated in Fig.~\ref{fig4} where $P(s)$ is 
obtained from the analytical eigenvalues of the free 
Dirac operator on a $53\times 47\times 43 \times 41$ lattice:
\begin{equation}
\label{free}
a^2\lambda^2 = \sum\limits_{\mu=1}^4 \sin^2 \left( \frac{2\pi n_{\mu}}{L_{\mu}}
\right)\:.
\end{equation}
Here $a$ is the lattice constant, $L_{\mu}$ is the number of lattice sites
in $\mu$-direction, and $n_{\mu}=0,...,L_{\mu}-1$. We used an asymmetric 
lattice with $L_{\mu}$ being primes and restricted the range to
$(L_{\mu}-1)/2$ instead of $L_{\mu}-1$ in each direction to avoid 
degeneracies of the free spectrum \cite{Verb}.

\section{Conclusion}

We have analyzed the nearest-neighbor spacing distribution
$P(s)$ of the eigenvalues of the Dirac operator in quenched $4d$ QED
on $8^3 \times 4$  and $8^3 \times 6$ lattices  both in the
strong coupling region and in the Coulomb phase. In both phases
we found excellent agreement of the lattice data with the Wigner surmise
of the unitary ensemble of RMT. Our results evidence that the 
fermions in U(1) gauge theory show quantum chaos in the
confined as well as in the Coulomb phase. Dynamical fermions are 
not expected to affect the Wigner distribution as has been demonstrated
for SU(3) \cite{Pull98}. In accordance with 
previous findings \cite{Pull98,Verb95} we conjecture that
the eigenvalues of the Dirac operator of interacting quantum field
theories quite generally reveal chaos due to the disorder of 
the gauge field configurations. The free Dirac operator, in absence
of a covariant derivative and minimal gauge coupling, exhibits
regular behavior.

It would be interesting to study the relationship between
chaos to order transitions in classical 
\cite{Biro94,Biro98,Gatto98,Sala97,Mukk97,Mati97,Case98} and quantum field
theories. However, this faces several difficulties: The
available investigations of classical field theories focus
mainly on the gauge sector, whereas the numerical methods
employed here are only efficient for the fermion sector of
quantum field theory. A similar accurate determination
of the eigenvalue spectrum of the gauge sector necessitates
to construct the corresponding Fock space and to diagonalize
high-dimensional matrices which seems to be out of reach
for $4 d$ QED/QCD. On the other hand, for the classical limit
fermion sector studies of chaos have not yet been attempted.

%%%%%%%%%%%%%%%%%%%%%%%%%%%%%%%%%%%%%%%%%%%%%%%%%%%%%%%%%%%%%%%%%%%%%%%%%%%%%
\section*{Acknowledgments}
This research was partially funded by the Department of Energy under 
Contracts No. DE-FG02-97ER41022 and DE-FG05-85ER2500 and by the 
``Fonds zur F\"orderung der Wissenschaftlichen Forschung'' under Contract
P11456-PHY. We also acknowledge useful discussions with T.S.~Bir\'o,
K.~Rabitsch, and T.~Wettig.
%%%%%%%%%%%%%%%%%%%%%%%%%%%%%%%%%%%%%%%%%%%%%%%%%%%%%%%%%%%%%%%%%%%%%%%%%%%
%                      REFERENCES                                         %
%%%%%%%%%%%%%%%%%%%%%%%%%%%%%%%%%%%%%%%%%%%%%%%%%%%%%%%%%%%%%%%%%%%%%%%%%%%

%%%%%%%%%%%%%%%%%%%%%%%%%%%%%%%%%%%%%%%%%%%%%%%%%%%%%%%%%%%%%%%%%%%%%%%%%%%%

\end{document}